\documentclass[article]{aa}
\usepackage{graphicx}
\usepackage{txfonts}
\usepackage{multirow}
\usepackage{xcolor,colortbl}
\usepackage{subcaption}
\usepackage{adjustbox}
\usepackage[colorlinks=true,allcolors=blue]{hyperref}
\newcommand{\msun}{$\mathrm{M}_{\odot}$ }

\newcommand{\km}[1]{$\rm \ km\ s^{-1}$ {#1}}

\usepackage{amstext}

\begin{document} 
        
        \title{Dynamical signature of a stellar bulge \\ in a quasar-host galaxy at $z\simeq6$}
        
        \author{R. Tripodi
                \inst{1,2,3}
                \and F. Lelli
                \inst{4}
                \and C. Feruglio
                \inst{2,3}
                \and F. Fiore
                \inst{1,2,3}
                \and F. Fontanot
                \inst{2,3}
                \and M. Bischetti
                \inst{1,2}
                \and R. Maiolino
                \inst{5,6,7}
        }
        
        \institute{Dipartimento di Fisica, Università di Trieste, Sezione di Astronomia, Via G.B. Tiepolo 11, I-34131 Trieste, Italy 
                \and
                INAF - Osservatorio Astronomico di Trieste, Via G. Tiepolo 11, I-34143 Trieste, Italy \\
                \email{roberta.tripodi@inaf.it}
                \and
                IFPU - Institute for Fundamental Physics of the Universe, via Beirut 2, I-34151 Trieste, Italy
                \and
                INAF - Osservatorio Astrofisico di Arcetri, Largo E. Fermi 5, 50125, Firenze, Italy
                \and 
                Institute of Astronomy, University of Cambridge, Madingley Road, Cambridge CB3 0HA, UK
                \and
                Kavli Institute for Cosmology, University of Cambridge, Madingley Road, Cambridge CB3 0HA, UK
                \and 
                Department of Physics and Astronomy, University College London, Gower Street, London WC1E 6BT, UK
                \\
        }
        
        \date{}
        
        \abstract{We present a dynamical analysis of a quasar-host galaxy at $z\simeq 6$ (SDSS J2310+1855) using a high-resolution ALMA observation of the [CII] emission line. The observed rotation curve was fitted with mass models that considered the gravitational contribution of a thick gas disc, a thick star-forming stellar disc, and a central mass concentration, which is likely due to a combination of a spheroidal component (i.e. a stellar bulge) and a supermassive black hole (SMBH). The SMBH mass of $5\times 10^9\ \rm M_{\odot}$, previously measured using the CIV and MgII emission lines, is not sufficient to explain the high velocities in the central regions. Our dynamical model suggests the presence of a stellar bulge with a mass of $\rm M_{bulge}\sim 10^{10}$\msun in this object, when the Universe was younger than 1 Gyr. To finally be located on the local $M_{\rm SMBH}-M_{\rm bulge}$ relation, the bulge mass should increase by a factor of $\sim$40 from $z=6$ to 0, while the SMBH mass should grow by a factor of 4 at most. This points towards asynchronous galaxy-BH co-evolution. Imaging with the JWST will allow us to validate this scenario.}

\maketitle

\section{Introduction}
\label{sec:intro}

The formation of central mass concentrations in galaxies, loosely defined as ``bulges'', is still poorly understood. 
The main proposed scenarios for bulge formation involve major mergers, which produce spheroids with a large Sérsic index \citep{hernquist1991}, and disc instabilities such as the formation of thick, buckling bars, which produce profiles with a smaller Sérsic index (the so-called pseudo-bulges; see e.g. \citealt{combes1981}). The latter imply long timescales and therefore occur mainly at late cosmic times. At high redshift, disc instabilities manifest as turbulent discs hosting giant clumps that can coalesce to form bulges \citep{bournaud2007,elmegreen2008}. In addition, when a luminous active galactic nucleus (AGN) is present, its energy output in the form of radiation pressure on dust or shocks can contribute to the formation of a spheroidal component by triggering both galactic outflows and star formation (SF) \citep{ishibashi2014, maiolino2017}.

Recent observations of the [CII] line at 158$\mu$m with the Atacama Millimeter/sub-millimeter Array (ALMA) revealed that stellar bulges can be in place up to $z\simeq5$ \citep{lelli2021, rizzo2021}, when the Universe was only 1.2 Gyr old. This evidence is based on gas dynamics: the inner [CII] rotation velocities are very high ($V_{\rm rot}\simeq300-400$ \km) and show a Keplerian-like decline out to $R\simeq1-3$ kpc, which can only be explained by a compact mass concentration. Similar rotation curves are indeed routinely observed in bulge-dominated galaxies at $z\simeq0$ \citep[e.g.][]{casertano1991, noorder2007, lelli2016}. It is crucial to push these studies to even higher $z$ to establish when the first bulges are formed and to better understand the co-evolution (or lack of it) with their central supermassive black holes (SMBHs). 

\begin{figure*}
        \centering
        \includegraphics[width=0.9\linewidth]{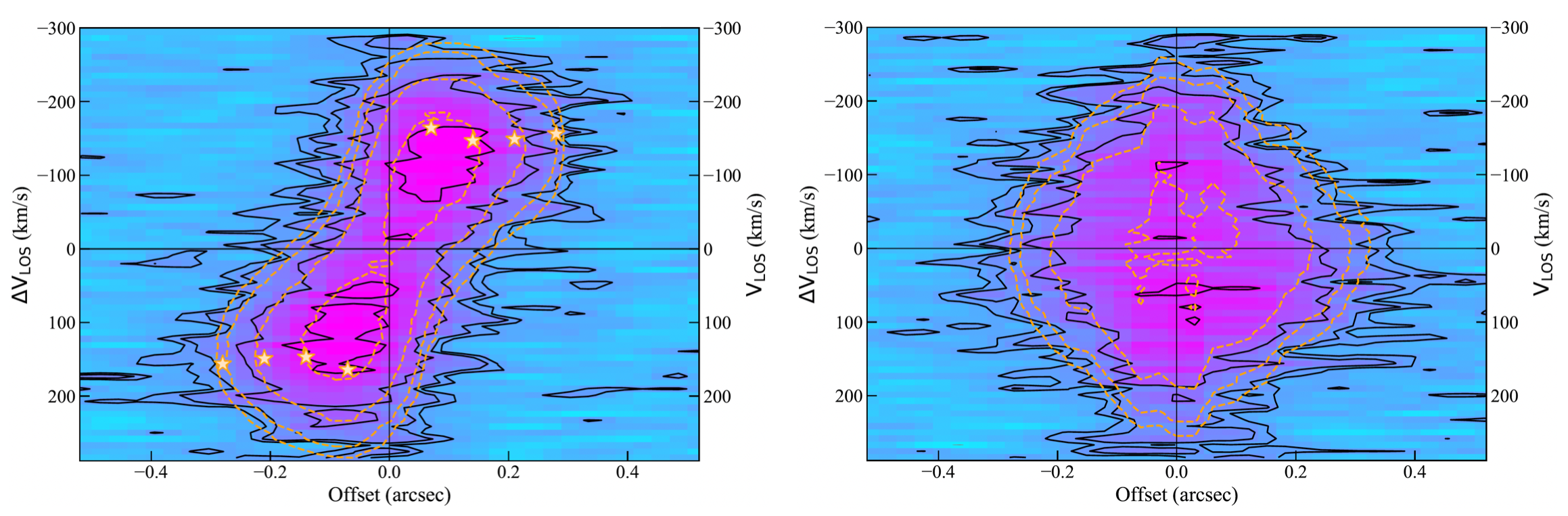}
        \caption{PV diagrams of the [CII] emission line along the kinematic major axis (left panel) and kinematic minor axis (right panel), performed with $^{\rm 3D}$Barolo. Contours are at $2,3,6,\text{and }12 \sigma$, with $\sigma = 0.22$ mJy/beam, for the data (solid black lines) and the best-fit model (dashed orange lines). Yellow stars show the projected rotation curve modelled by $^{\rm 3D}$Barolo.} 
        
        \label{fig:pvdiagram}
\end{figure*}

We performed a dynamical decomposition of the rotation curve of Sloan Digital Sky Survey (SDSS) J231038.88+185519.7 (hereafter J2310) at $z=6.0031$ \citep{Tripodi2022, wang2013}, based on high-resolution ALMA observations. J2310 was first identified as a quasi-stellar object (QSO) in SDSS \citep{jiang2006,wang2013} and is one of the most luminous QSOs known in the optical and far-infrared bands, with a bolometric luminosity of $L_{\rm bol}=9.3 \times 10^{13}\ \rm L_\odot$. 

\citet{bischetti2022} reported evidence for an AGN-driven ionised-gas wind traced by ultra-violet (UV) broad absorption lines. The host galaxy is characterised by a star formation rate (SFR) in the range $\sim$1200-2400 $\rm M_{\odot}\ yr^{-1}$ \citep{shao2019,Tripodi2022} that occurs in a regularly rotating gaseous disc without evidence of interactions or mergers.

\citet{Tripodi2022} presented a tilted-ring modelling of the [CII] emission that resulted in a flat rotation curve and derived a disc inclination $i \simeq 25^{\circ}$ and a dynamical mass $M_{\rm dyn}=5.2^{+2.3}_{-3.2}\times 10^{10}\ \rm M_{\odot}$. They interpreted the high velocities seen near the galactic centre as due to a [CII] outflow. An alternative possible explanation for the high-velocity [CII] emission is fast rotation due to the presence of a central mass concentration.
In this study, we investigate whether a bulge can indeed be the most likely cause for the velocity enhancement in the innermost region of J2310 by modelling the rotation curve with different mass models.

We adopted a $\Lambda$CDM cosmology from \citet{planck2016}: $H_0=67.7\ \rm km\ s^{-1}\ Mpc^{-1}$, $\Omega_m = 0.308$ and $\Omega_{\Lambda} = 0.692$. Thus, the angular scale is $5.84$ kpc/arcsec at $z=6$. We analysed the same [CII] cube as described in \citet{Tripodi2022}. It has a spatial resolution of ($0.17 \times 0.15$) arcsec$^2$, which corresponds to $\sim$0.9 kpc at the redshift of the QSO.

\section{Analysis and results}
\label{sec:analysis}
\subsection{[CII] rotation curve}
\label{sec:kin}

We derived a new rotation curve of J2310 using the software $^{\rm 3D}$Barolo \citep{diteodoro2015}, improving the preliminary kinematic modelling of \citet{Tripodi2022}. $^{\rm 3D}$Barolo directly fits disc models to the 3D cube and considers the effects of spectral and spatial resolution (beam smearing). The galaxy is divided into a series of rings, each of which is described by nine parameters: coordinates of the kinematic centre $(x_0,y_0)$, position angle (PA), inclination ($i$), vertical thickness ($z_0$), systemic velocity ($V_{\rm sys}$), rotation velocity ($V_{\rm rot}$), radial velocity ($V_{\rm rad}$), and velocity dispersion ($\sigma_v$). For $(x_0,y_0)$, $V_{\rm sys}$ , and $i$, we adopted the same estimates as in \citet{Tripodi2022}. We fixed $z_0$ to 300 pc, but this parameter has a negligible effect on our results because  our spatial resolution is $\sim$900 pc and the disc is nearly face-on ($i\simeq25^{\circ}$). Since the kinematic major axis is perpendicular to the minor axis, we fixed the radial velocity to zero. With respect to \citet{Tripodi2022}, we changed the PA to 195$^{\circ}$ because the data-model residual maps in their Figure 8 suggest that their value of 200$^{\circ}$ was slightly overestimated \citep[see][]{warner1973}. We also changed the radial sampling to avoid aggressive over-sampling; we used four rings spaced by 0.07 arcsec, which corresponds to 0.409 kpc. In order to minimise the impact of the outflowing material on our results, we cropped the [CII] cube considering the velocity range [-290, +270]\km. This excludes the blue and red wings seen in the spatially integrated [CII] spectrum (see Figure 2 in \citealt{Tripodi2022}). We then ran $^{\rm 3D}$Barolo with $v_{\rm rot}$ and $\sigma_v$ as free parameters. $^{\rm 3D}$Barolo provides asymmetric uncertainties ($\delta_+, \delta_{-}$), corresponding to a variation of 5\% in the residual from the global minimum. We treated them as $1\sigma$ uncertainties and took the mean value of ($\delta_+, \delta_{-}$) to compute symmetric uncertainties that were used in subsequent rotation-curve fits.

Figure \ref{fig:pvdiagram} shows position-velocity (PV) diagrams along the major and minor axes of the disc. In \citet{Tripodi2022}, the comparison between disc model and observations showed a mismatch at line-of-sight velocities $V_{\rm LOS}$ of $\pm$300\km along both the major and minor axes. This mismatch was interpreted as due to the gas outflow. The overall symmetry of the PV diagrams, however, may suggest that the high-velocity emission is associated with fast rotation. With the new rotation curve, the central mismatch at $\gtrsim 300$\km has now decreased along both axes, indicating that the new model fits the data better. 

The rotation curve shown in Figure \ref{fig:pvdiagram} is approximately flat, with a slight increase towards the centre. The average deprojected rotation velocity is $\sim$365 \km, similar to what was found in \citet{Tripodi2022}. The velocity dispersion profile (not shown) increases towards the centre and reaches $\sigma_{\rm gas}\simeq 80$ \km , with an average value of $\sim$70 \km. This gives $V_{\rm rot}/\sigma_{\rm gas}\simeq5$, so that the disc is rotation supported and the observed rotation velocities are a good tracer of the circular velocity of a test particle in the equilibrium gravitational potential. The circular velocity approximately goes as $(V_{\rm rot}^2 + \sigma_{\rm gas}^2)^{0.5}$, therefore corrections for pressure support are a few \km at most and are therefore negligible.

\begin{table}
        \caption{MCMC priors}
        \centering
        \begin{tabular}{c|ccc}
                \hline
                Parameter & Prior Type & Center & Standard Deviation  \\
                \hline
                $i$ [$^{\circ}$] & G & 25 & 5 \\
                $\log(Y_{\rm gas})$ & LN & 0.64 & 0.11 \\
                $\log(Y_{\rm *disc})$ & LN & 0.0 & 1.0 \\
                $\log(Y_{\rm bulge})$ & LN & 0.0 & 1.0\\
                $\log(Y_{\rm BH})$ & LN & 0.0 & 1.0\\
                $M_{\rm BH}$ [$\rm M_{\odot}$] & Fixed & $5\times10^{9}$ & ... \\
                \hline
        \end{tabular}
        \label{tab:priors}
        \flushleft 
        \footnotesize {{\bf Notes.} G = Gaussian. LN = Lognormal. A prior for $\log(Y_j)$ of $0.0\pm1.0$ corresponds to $(1\pm10)\times 10^{10}$\msun.}
\end{table}

\begin{figure*}
        \centering
        \includegraphics[width=0.9\linewidth]{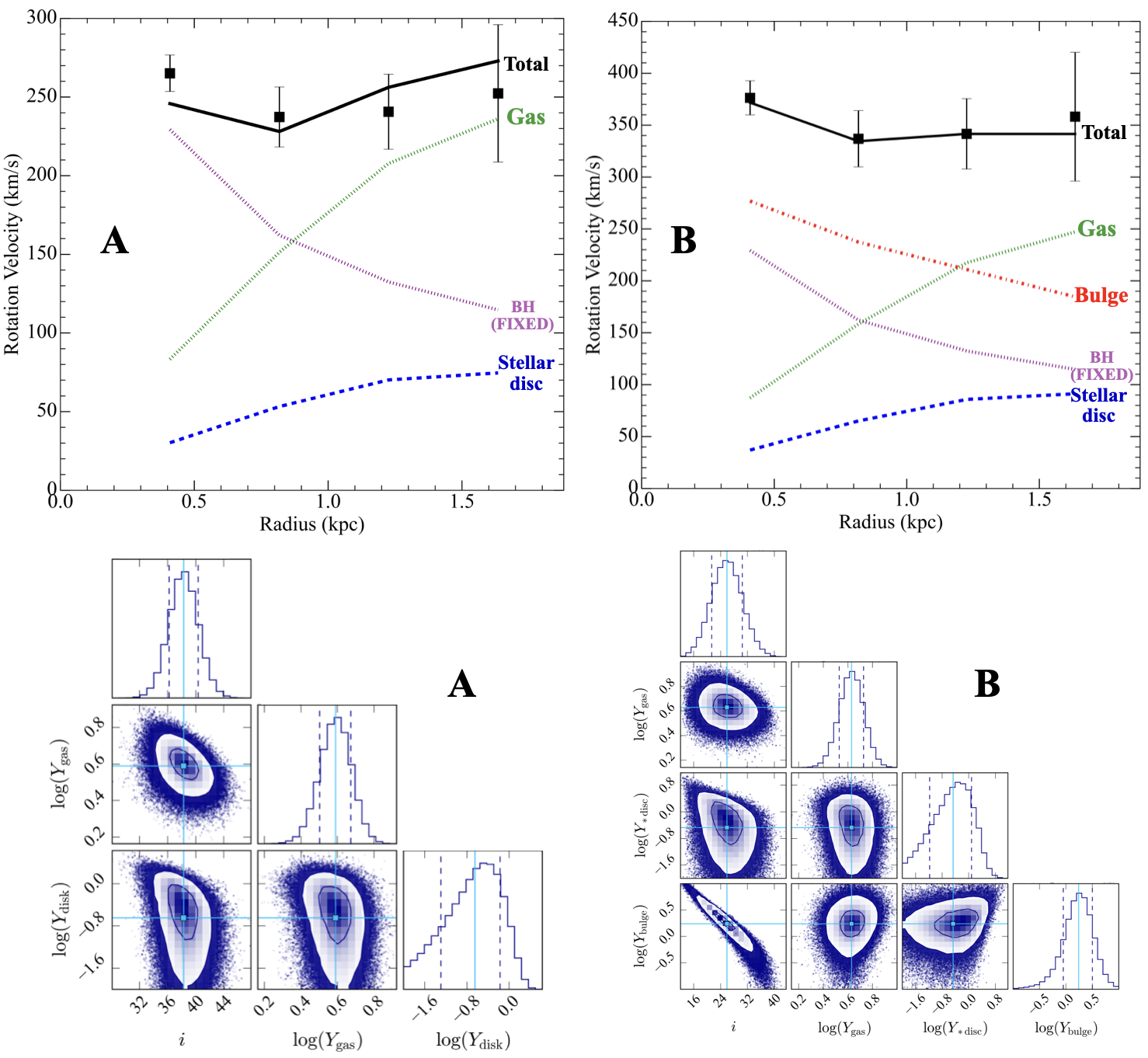}
        \caption{Results for mass models. Top panels: Two different mass models are fitted to the observed rotation curve (black squares): no bulge and fixed BH mass (A),  bulge and fixed BH mass (B). The best-fitting model (solid black line) is the sum of the contributions from the stellar disc (dashed blue line), the cold gas disc (dotted green line), the BH (dotted purple line), and the stellar bulge (when present; dot-dashed red line). The different values of the rotation velocities in models A and B are due to different best-fitting values for the inclination: $i=38 \pm 2^{\circ}$ for model A and $i=26 \pm 4^{\circ}$ for model B. Bottom panels: Posterior distributions of the fitting parameters for models A and B.}
        \label{fig:models}
\end{figure*}

\subsection{Mass models}
\label{sec:model}

We used the observed rotation curve to constrain the mass distribution within the galaxy. We considered three to four mass contributions: a gaseous disc, a stellar disc, an SMBH, and possibly, a stellar bulge. The model circular velocity at radius $R$ is then given by
\begin{equation}
        \label{eq:vrot}
        V_{\rm mod}^2 = Y_{\rm gas} V^2_{\rm gas}+Y_{\rm *disc}V^2_{\rm *disc}+Y_{\rm bulge}V^2_{\rm bulge}+V^2_{\rm BH}
,\end{equation}
where $Y_j=\dfrac{M_j}{10^{10}\ \rm M_{\odot}}$ are dimensionless parameters of the order of one (with j = gas, $*$disc, bulge). Essentially, for numerical convenience, we normalised the gravitational contributions to a mass of 10$^{10}$ M$_\odot$ and scaled them using $Y_{j}$. The gaseous and stellar disc were both modelled as thick discs with a radial density profile given by the observed [CII] and dust surface brightness profiles, respectively (following \citealt{lelli2021}). For the vertical density profile, we assumed an exponential function with a fixed scale height. The stellar bulge was modelled as a spherical projected De Vaucouleurs profile with a fixed effective radius of 0.07 arcsec ($\sim$409 pc). This is the maximum effective radius allowed by the data, so that our bulge model maximises the difference between an extended mass distribution and a point source. The SMBH was modelled as a point mass. A detailed description of our assumptions is given in Appendix \ref{sec:models-details}. 

The masses of the stellar and gas discs are degenerate because their velocity contributions have a similar shape, which rises with radius. To break this degeneracy, we imposed a tight prior on $Y_{\rm gas}$ by adopting a gas mass of $4.4\times 10^{10}$ \msun  \citep[][for more details see Appendix \ref{sec:models-details}]{feruglio2018, li2020, Tripodi2022}. Similarly, the masses of bulge and SMBH are degenerate because their velocity contributions declines with radius. This occurs because the spatial resolution of the rotation curve cannot discern a central point source from a more extended central mass concentration. To break this degeneracy, we fixed the SMBH mass to $5 \times 10^{9}$ \msun (Mazzucchelli et al. in prep.; \citealt{bischetti2022}). This estimate comes from X-SHOOTER spectroscopy of the MgII and CIV lines, which has a high signal-to-noise ratio and assumes the $z\simeq0$ scaling relations provided by \citet{vosmer2009}, \citet{vestergard2006}, and \citet{coatman2017}.

Depending on the chosen model, we therefore had three or four free parameters: the inclination $i$ of the [CII] disc that changes the observed rotation velocities as $V_{\rm obs}\sin(i)$, and two or three parameters among $Y_{\rm gas}$, $Y_{\rm *disc}$, and $Y_{\rm bulge}$. The best-fitting parameters were determined using a Markov chain Monte Carlo (MCMC) method in a Bayesian context. We adopted a Gaussian prior on $i$ and lognormal priors on the $Y_j$ parameters. The choices for the central and standard deviation values for all the fitting parameters are listed in Table \ref{tab:priors}.

\begin{table}
        \caption{MCMC results for mass model B.}
        \centering
        \begin{tabular}{c|cc}
                \hline
                Parameter & Best-fitting values  \\
                \hline
                $i$ [$^{\circ}$] &  $26\pm4$\\[0.1cm]
                $M_{\rm gas} [10^{10}\ \rm M_{\odot}]$ & $4.26\pm0.97$\\[0.1cm]
                $M_{\rm *disc} [10^{10}\ \rm M_{\odot}]$ & $0.34^{+0.43}_{-0.55}$\\[0.1cm]
                $M_{\rm bulge} [10^{10}\ \rm M_{\odot}]$ & $1.74^{+1.04}_{-1.17}$\\[0.1cm]
                $M_{\rm BH} [10^{10}\ \rm M_{\odot}]$  &  $0.5$ (Fixed) \\ [0.1cm]
                \hline
        \end{tabular}
        \label{tab:results}
        \flushleft 
\end{table}

To assess whether the bulge is a truly essential component, we started with a conservative model that considered only stellar disc, gas disc, and SMBH with a fixed mass. Figure \ref{fig:models} (model A, left panel) shows that this model misses the first point of the rotation, and the inclination is pushed up to 38$^{\circ}$ (more than 2.5$\sigma$ from the prior value) to decrease the rotation velocities as much as possible. When we instead consider a model in which the BH mass is entirely free, we obtain a good result, but the required BH mass is higher than the expected value derived from the MgII line by a factor of 2 (i.e. $M_{\rm BH}=1.1\pm 0.4 \times 10^{10}\ \rm M_{\odot}$). This high mass is still within the systematic uncertainties of $\sim$0.3 dex associated with BH masses from MgII measurements \citep{vestergard2006}. An increase in the BH mass of a factor of 2 would therefore be able to remove the need for an additional central stellar concentration. However, this scenario seems very unlikely according to galaxy formation models (see Sect.\,\ref{sec:disc} and Figure\,\ref{fig:mbh-msph}). 

Finally, we considered a mass model that included both an SMBH with a fixed mass and a bulge component. Figure \ref{fig:models} (model B, right panel) shows that this model reproduces the observed rotation curve well and preserves the prior value of $i\simeq25^{\circ}$. The best-fitting values are reported in Table \ref{tab:results}. We find a bulge mass of $\sim 1.7\times 10^{10}$\msun that implies $M_{\rm bulge}\simeq 0.27 \times M_{\rm baryon}$, where $M_{\rm baryon}=M_{\rm gas}+M_{\rm *disc}+M_{\rm bulge}$. The best-fit mass also implies an effective surface mass density $\Sigma_{\rm eff}\simeq3\times10^4$ M$_\odot$ pc$^{-2}$ , which is consistent with that of the most compact spheroids at $z\simeq0$. It seems unlikely that such a compact mass component can be in the form of [CII]-dark or CO-dark gas because it should lead to a strong central enhancement of star formation that is not visible in the sub-millimeter continuum map. At larger radii, however, the gas contribution becomes strong, and the galaxy has $M_{\rm gas}\simeq 0.67 \times M_{\rm baryon}$; this result is largely driven by the tight prior on the gas mass.

Recently, \citet{shao2022} presented a similar analysis of the rotation curve of J2310, using a razor-thin gas disc and a single spherical stellar component. They obtained a stellar mass of $5.8\times 10^9\ \rm M_{\odot}$ when they fixed the BH mass to the value of $1.8\times 10^9$ that was derived from the analysis of the CIV emission line \citep{feruglio2018}. Our mass models are different from theirs because we distinguish between a star-forming stellar disc and a spherical bulge, which is crucial to locate J2310 on the BH-bulge scaling relation (see Sect. \ref{sec:disc}). In addition, in our models, the gas and stellar discs have a realistic finite thickness. Importantly, \citet{shao2022} concluded that they measured the dynamical mass of a black hole for the first time at $z\simeq6$, whereas we find that the BH mass and bulge mass are degenerate. To break this degeneracy, the BH mass must be fixed using independent information, given the low spatial resolution of the ALMA data ($\sim$900 pc). For comparison, dynamical BH mass measurements in AGN-host galaxies at $z\simeq0$ require spatial resolutions better than 100 pc \citep[e.g.][]{combes2019, lelli2022}. The observed rotation curve of J2310 clearly indicates a central mass concentration, but the ALMA data themselves cannot determine whether this is entirely due to a BH, a stellar bulge, or a combination of both. In our view, the latter scenario is the most likely.

\section{Discussion and conclusions}
\label{sec:disc}

Our dynamical analysis suggests the presence of a central compact mass component in J2310. This component is composed of an SMBH and a stellar bulge with similar masses of the order of $\sim 1.7 \times 10^{10}$ M$_\odot$. Two crucial questions that arise from this result are first, how a bulge of $\sim 10^{10}$\msun was able to form in less than 1Gyr, and second, what the implications of our results are for the overall picture of SMBH-galaxy co-evolution.

As we discussed in Sect.\,\ref{sec:intro}, several mechanisms have been proposed for the formation of galaxy bulges: disc instabilities \citep{combes1981, bournaud2007}, major mergers \citep{toomre1977a,hernquist1991}, SF in AGN driven outflows \citep{ishibashi2014}, and direct dissipative collapse \citep{Eggen1962, sandage1990}. For J2310, disc instabilities can probably be ruled out because these secular processes occur on long timescales ($\sim$3-5 Gyr), while the age of the Universe at $z\simeq6$ is only $\sim$900 Myr in the adopted cosmology. All the other mechanisms appear viable, however. The host galaxy of J2310 does not show any evidence of ongoing mergers based on the high-resolution ALMA data that are available to date, but the orbital time at the last measured point of the rotation curve is only 26 Myr, which means that the [CII] disc may relax in just $\sim$100 Myr ($\text{about four}$ revolutions) after a violent event. This implies that the last potential merger between (proto-)galaxies with masses of a few times 10$^9$ M$_\odot$ must have occurred at $z>6.5$.

Feedback from AGN in the form of large-scale outflows driven by the radiation pressure on dust or by shock propagation may trigger the formation of new stars at outer radii, leading to the development of extended stellar envelopes  \citep{ishibashi2014}. Observational evidence supporting this scenario was provided by \citet{maiolino2017}, who found that SF can occur within galactic outflows where the newly formed stars follow ballistic trajectories that can fall back in the potential and thus form a spheroidal component. In this scenario, the accreting BH triggers SF and contributes to the formation of the stellar bulge. This scenario may form extended and diffuse bulges, while in J2310, we find evidence for a compact mass concentration with a high effective surface density.

\begin{figure}
        \centering
        \includegraphics[width=0.9\linewidth]{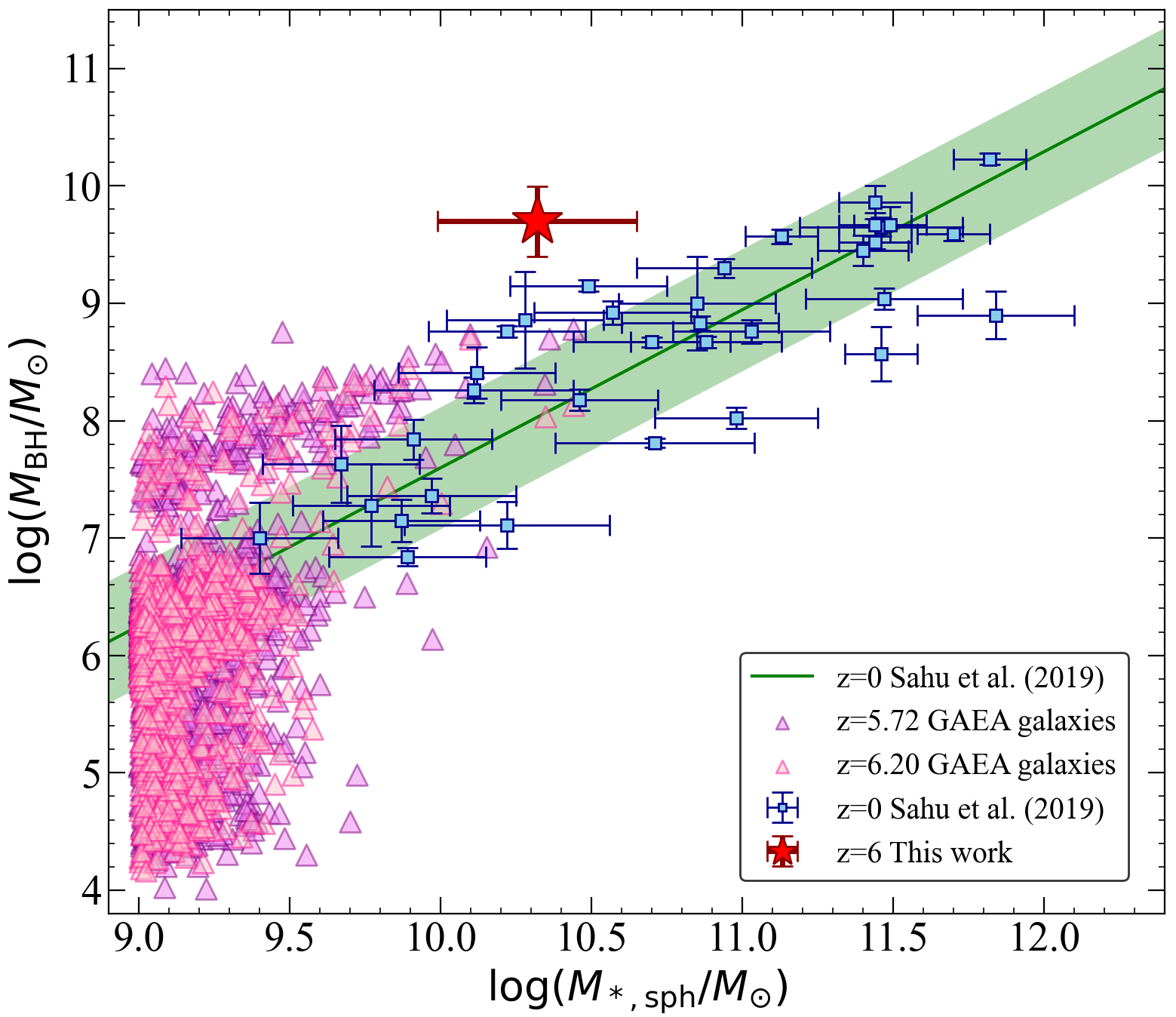}
        \caption{Black hole mass vs. spheroid stellar mass. Comparison between the $z=0$ sample of early-type galaxies from \citet{sahu2019} (blue squares) and our results for the $z=6$ QSO J2310 (red star). The green line is the best-fit relation for the $z=0$ sample. Its scatter (green shadowed region) is taken from \citet{sahu2019}. The purple and pink triangles are the $z=5.72$ and $z=6.20$ galaxies, respectively, that were modelled by the GAEA-F06 run \citep{fontanot2020}.}
        \label{fig:mbh-msph}
\end{figure}

Regarding the BH-galaxy co-evolution, for J2310, \citet{Tripodi2022} found that the SMBH growth efficiency is $\sim$50\% lower than that of its host galaxy. This suggests that AGN feedback effectively slows down the accretion onto the SMBH, while the host galaxy is still growing fast. The onset of significant BH feedback hampering BH growth marks the transition from a phase of BH dominance to a phase of symbiotic growth of the BH and the galaxy \citep{volonteri2012}. In Figure \ref{fig:mbh-msph}, we locate J2310 on the local $M_{\rm BH}-M_{\rm bulge}$ relation (green region) and the local observational data (blue squares) from \citet{sahu2019}. J2310 deviates from the $z\simeq0$ relation because it has $M_{\rm BH}\simeq 0.3 M_{\rm bulge}$  , while in local galaxies, $M_{\rm BH}\simeq (0.01-0.1) \times M_{\rm bulge}$. When we assume that the BH of J2310 evolves into one of the most massive BH at $z=0$ ($M_{\rm BH}\simeq 2\times 10^{10}$ \msun), it can grow in mass by only a factor of 4 during $\sim$13 Gyr, while the bulge should grow by a factor of $\sim$40 to reach a mass around $ 10^{11.8}$\msun at $z\simeq0$. This suggests an asynchronous growth between SMBH and bulges. Moreover, the available molecular gas in J3210 is estimated to be $\sim 4\times10^{10}\ \rm M_\odot$. To reach a stellar mass of $\sim 10^{11.8}$ M$_\odot$ at $z\simeq0$, the galaxy needs to accrete a comparable amount of mass (in gas or directly in stars) over $\sim$13 Gyr, giving an average mass accretion rate of $\sim$ 40 M$_\odot$ yr$^{-1}$. For example, if the galaxy continues to form stars with  $\rm{SFR}\simeq10^{3}$ M$_\odot$ yr$^{-1}$, a stellar mass of $(5-10)\times10^{11}$ M$_\odot$ can be formed in only 0.5-1.0 Gyr (by $z\simeq3.5-4.0$), but would then require a mass-accretion rate comparable to the SFR. In any case, our result seems to highlight the absence of any symbiotic growth between the SMBH and the host galaxy at high redshift; it appears that the SMBH firstly experiences a phase of rapid and intense growth (reaching up to $5\times 10^9$\msun in less than 1 Gyr), followed by the growth of the galaxy.

Finally, we compared our result with theoretical predictions of the GAlaxy Evolution and Assembly (GAEA) model (in detail the GAEA-F06 run, \citealt{fontanot2020}), to shed light on the mechanism leading to the formation of this bulge.
 This semi-analytic model includes two different channels of bulge formation (i.e. galaxy mergers and disc instabilities; see e.g. \citealt{delucia2011}): both physical processes also help bringing cold gas onto the central SMBH, thus feeding accretion and AGN events. In Figure \ref{fig:mbh-msph} we also show the distribution of model galaxies\footnote{The GAEA-F06 realisation has been defined on merger trees extracted from the Millennium Simulation. We considered model galaxies in the two snapshots closer to the estimated J2310 redshift, $z=6.20$ and $z=5.72$.} in the $M_{\rm BH}-M_{\rm bulge}$ plane (purple and pink triangles) compared with a local $z\sim0$ sample \citep{sahu2019}. From the comparison, we can conclude that (a) GAEA-F06 predicts a relevant population of $M_{\rm bulge}>10^9$\msun  bulges in high-z galaxies despite the relatively young cosmic epoch, and (b) the scatter around the $z\sim6$ $M_{\rm BH}-M_{\rm bulge}$ relation is much larger than in the local Universe. This is likely due to the different contribution of the two bulge-forming channels (mergers and disc instabilities) to the corresponding SMBH accretion \citep[see][]{fontanot2020}: in this interpretation, at fixed $M_{\rm bulge}$, higher (lower) $M_{\rm BH}$ corresponds to sources that experienced more mergers (disc instabilities). We plan to further test this hypothesis in forthcoming work. We verified that GAEA predicts that all objects lying above the $M_{\rm BH}-M_{\rm bulge}$ relation at $z\sim6$ converge towards the $z\sim0$ relation at later times. Nonetheless, none of the model galaxies reaches $z\sim6 \ M_{\rm BH}$ as high as the one estimated for J2310, which thus represents the main difference between data and model predictions.

A robust interpretation of our results is strongly dependent on the accuracy of the determination of the physical quantities at play. In particular, to be able to precisely quantify the mass of the central bulge, we would need a [CII] observation with higher resolution. To verify this, we simulated a rotation curve derived from an observation with a resolution of $\sim 0.05$ arcsec, that is, an improvement of a factor of $\times3$ (linear) compared to  our current resolution, and we find that this  observation would allow us to constrain the bulge mass to within an uncertainty of 30\% (for further details, see Appendix \ref{app:simul}). This observation would be feasible with ALMA.

\begin{acknowledgements}
         We thank the referee for the useful suggestions. RT thanks Marta Molero for the insightful discussions. This paper makes use of the following ALMA data:  ADS/JAO.ALMA\#2019.1.00661.S. ALMA is a partnership of ESO (representing its member states), NFS (USA) and NINS (Japan), together with NRC (Canada), MOST and ASIAA (Taiwan) and KASI (Republic of Korea), in cooperation with the Republic of Chile. The Joint ALMA Observatory is operated by ESO, AUI/NRAO and NAOJ. RT acknowledges financial support from the University of Trieste. RT, CF, FF, MB acknowledge support from PRIN MIUR project “Black Hole winds and the Baryon Life Cycle of Galaxies: the stone-guest at the galaxy evolution supper”, contract \#2017PH3WAT. RM acknowledges ERC Advanced Grant 695671 QUENCH, and support from the UK Science and Technology Facilities Council (STFC). RM also acknowledges funding from a research professorship from the Royal Society. The authors thank Pierluigi Monaco for his helpful suggestions. This paper makes extensive use of \textit{python} packages, libraries and routines, such as \texttt{numpy}, \texttt{scipy} and \texttt{astropy}.
        \textit{Facilities:} ALMA. \textit{Software:} CASA (v5.1.1-5, \citealt{mcmullin2007}), $^{\rm 3D}$Barolo \citep{diteodoro2015}. 
        
\end{acknowledgements}

\bibliographystyle{aa} 
\bibliography{biblio}

\begin{thebibliography}{47}
\expandafter\ifx\csname natexlab\endcsname\relax\def\natexlab#1{#1}\fi

\bibitem[{{Accurso} {et~al.}(2017){Accurso}, {Saintonge}, {Catinella},
  {Cortese}, {Dav{\'e}}, {Dunsheath}, {Genzel}, {Gracia-Carpio}, {Heckman},
  {Jimmy}, {Kramer}, {Li}, {Lutz}, {Schiminovich}, {Schuster}, {Sternberg},
  {Sturm}, {Tacconi}, {Tran}, \& {Wang}}]{accurso2017}
{Accurso}, G., {Saintonge}, A., {Catinella}, B., {et~al.} 2017, \mnras, 470,
  4750

\bibitem[{{Bershady} {et~al.}(2010){Bershady}, {Verheijen}, {Swaters},
  {Andersen}, {Westfall}, \& {Martinsson}}]{bershady2010}
{Bershady}, M.~A., {Verheijen}, M. A.~W., {Swaters}, R.~A., {et~al.} 2010,
  \apj, 716, 198

\bibitem[{{Bischetti} {et~al.}(2022){Bischetti}, {Feruglio}, {D'Odorico},
  {Arav}, {Ba{\~n}ados}, {Becker}, {Bosman}, {Carniani}, {Cristiani}, {Cupani},
  {Davies}, {Eilers}, {Farina}, {Ferrara}, {Maiolino}, {Mazzucchelli},
  {Mesinger}, {Meyer}, {Onoue}, {Piconcelli}, {Ryan-Weber}, {Schindler},
  {Wang}, {Yang}, {Zhu}, \& {Fiore}}]{bischetti2022}
{Bischetti}, M., {Feruglio}, C., {D'Odorico}, V., {et~al.} 2022, \nat, 605, 244

\bibitem[{{Bolatto} {et~al.}(2013){Bolatto}, {Wolfire}, \&
  {Leroy}}]{Bolatto2013}
{Bolatto}, A.~D., {Wolfire}, M., \& {Leroy}, A.~K. 2013, \araa, 51, 207

\bibitem[{{Bournaud} {et~al.}(2007){Bournaud}, {Elmegreen}, \&
  {Elmegreen}}]{bournaud2007}
{Bournaud}, F., {Elmegreen}, B.~G., \& {Elmegreen}, D.~M. 2007, \apj, 670, 237

\bibitem[{{Carilli} \& {Walter}(2013)}]{carilli2013}
{Carilli}, C.~L. \& {Walter}, F. 2013, \araa, 51, 105

\bibitem[{{Casertano}(1983)}]{casertano1983}
{Casertano}, S. 1983, \mnras, 203, 735

\bibitem[{{Casertano} \& {van Gorkom}(1991)}]{casertano1991}
{Casertano}, S. \& {van Gorkom}, J.~H. 1991, \aj, 101, 1231

\bibitem[{{Coatman} {et~al.}(2017){Coatman}, {Hewett}, {Banerji}, {Richards},
  {Hennawi}, \& {Prochaska}}]{coatman2017}
{Coatman}, L., {Hewett}, P.~C., {Banerji}, M., {et~al.} 2017, \mnras, 465, 2120

\bibitem[{{Combes} {et~al.}(2019){Combes}, {Garc{\'\i}a-Burillo}, {Audibert},
  {Hunt}, {Eckart}, {Aalto}, {Casasola}, {Boone}, {Krips}, {Viti}, {Sakamoto},
  {Muller}, {Dasyra}, {van der Werf}, \& {Martin}}]{combes2019}
{Combes}, F., {Garc{\'\i}a-Burillo}, S., {Audibert}, A., {et~al.} 2019, \aap,
  623, A79

\bibitem[{{Combes} \& {Sanders}(1981)}]{combes1981}
{Combes}, F. \& {Sanders}, R.~H. 1981, \aap, 96, 164

\bibitem[{{De Lucia} {et~al.}(2011){De Lucia}, {Fontanot}, {Wilman}, \&
  {Monaco}}]{delucia2011}
{De Lucia}, G., {Fontanot}, F., {Wilman}, D., \& {Monaco}, P. 2011, \mnras,
  414, 1439

\bibitem[{{Di Teodoro} \& {Fraternali}(2015)}]{diteodoro2015}
{Di Teodoro}, E.~M. \& {Fraternali}, F. 2015, \mnras, 451, 3021

\bibitem[{{Eggen} {et~al.}(1962){Eggen}, {Lynden-Bell}, \&
  {Sandage}}]{Eggen1962}
{Eggen}, O.~J., {Lynden-Bell}, D., \& {Sandage}, A.~R. 1962, \apj, 136, 748

\bibitem[{{Elmegreen} {et~al.}(2008){Elmegreen}, {Bournaud}, \&
  {Elmegreen}}]{elmegreen2008}
{Elmegreen}, B.~G., {Bournaud}, F., \& {Elmegreen}, D.~M. 2008, \apj, 688, 67

\bibitem[{{Feruglio} {et~al.}(2018){Feruglio}, {Fiore}, {Carniani}, {Maiolino},
  {D'Odorico}, {Luminari}, {Barai}, {Bischetti}, {Bongiorno}, {Cristiani},
  {Ferrara}, {Gallerani}, {Marconi}, {Pallottini}, {Piconcelli}, \&
  {Zappacosta}}]{feruglio2018}
{Feruglio}, C., {Fiore}, F., {Carniani}, S., {et~al.} 2018, \aap, 619, A39

\bibitem[{{Fontanot} {et~al.}(2020){Fontanot}, {De Lucia}, {Hirschmann}, {Xie},
  {Monaco}, {Menci}, {Fiore}, {Feruglio}, {Cristiani}, \&
  {Shankar}}]{fontanot2020}
{Fontanot}, F., {De Lucia}, G., {Hirschmann}, M., {et~al.} 2020, \mnras, 496,
  3943

\bibitem[{{Genzel}(2017)}]{genzel2017}
{Genzel}, R. 2017, in Disk Instabilities Across Cosmic Scales, 5

\bibitem[{{Genzel} {et~al.}(2020){Genzel}, {Price}, {{\"U}bler}, {F{\"o}rster
  Schreiber}, {Shimizu}, {Tacconi}, {Bender}, {Burkert}, {Contursi}, {Coogan},
  {Davies}, {Davies}, {Dekel}, {Herrera-Camus}, {Lee}, {Lutz}, {Naab}, {Neri},
  {Nestor}, {Renzini}, {Saglia}, {Schuster}, {Sternberg}, {Wisnioski}, \&
  {Wuyts}}]{genzel2020}
{Genzel}, R., {Price}, S.~H., {{\"U}bler}, H., {et~al.} 2020, \apj, 902, 98

\bibitem[{{Hernquist} \& {Barnes}(1991)}]{hernquist1991}
{Hernquist}, L. \& {Barnes}, J.~E. 1991, \nat, 354, 210

\bibitem[{{Ishibashi} \& {Fabian}(2014)}]{ishibashi2014}
{Ishibashi}, W. \& {Fabian}, A.~C. 2014, \mnras, 441, 1474

\bibitem[{{Jiang} {et~al.}(2006){Jiang}, {Fan}, {Hines}, {Shi}, {Vestergaard},
  {Bertoldi}, {Brandt}, {Carilli}, {Cox}, {Le Floc'h}, {Pentericci},
  {Richards}, {Rieke}, {Schneider}, {Strauss}, {Walter}, \&
  {Brinkmann}}]{jiang2006}
{Jiang}, L., {Fan}, X., {Hines}, D.~C., {et~al.} 2006, \aj, 132, 2127

\bibitem[{{Lelli} {et~al.}(2022){Lelli}, {Davis}, {Bureau}, {Cappellari},
  {Liu}, {Ruffa}, {Smith}, \& {Williams}}]{lelli2022}
{Lelli}, F., {Davis}, T.~A., {Bureau}, M., {et~al.} 2022, \mnras, 516, 4066

\bibitem[{{Lelli} {et~al.}(2021){Lelli}, {Di Teodoro}, {Fraternali}, {Man},
  {Zhang}, {De Breuck}, {Davis}, \& {Maiolino}}]{lelli2021}
{Lelli}, F., {Di Teodoro}, E.~M., {Fraternali}, F., {et~al.} 2021, Science,
  371, 713

\bibitem[{{Lelli} {et~al.}(2016){Lelli}, {McGaugh}, \& {Schombert}}]{lelli2016}
{Lelli}, F., {McGaugh}, S.~S., \& {Schombert}, J.~M. 2016, \aj, 152, 157

\bibitem[{{Li} {et~al.}(2020){Li}, {Wang}, {Riechers}, {Walter}, {Decarli},
  {Venamans}, {Neri}, {Shao}, {Fan}, {Gao}, {Carilli}, {Omont}, {Cox},
  {Menten}, {Wagg}, {Bertoldi}, \& {Narayanan}}]{li2020}
{Li}, J., {Wang}, R., {Riechers}, D., {et~al.} 2020, \apj, 889, 162

\bibitem[{{Lima Neto} {et~al.}(1999){Lima Neto}, {Gerbal}, \&
  {M{\'a}rquez}}]{lima1999}
{Lima Neto}, G.~B., {Gerbal}, D., \& {M{\'a}rquez}, I. 1999, \mnras, 309, 481

\bibitem[{{Maiolino} {et~al.}(2017){Maiolino}, {Russell}, {Fabian}, {Carniani},
  {Gallagher}, {Cazzoli}, {Arribas}, {Belfiore}, {Bellocchi}, {Colina},
  {Cresci}, {Ishibashi}, {Marconi}, {Mannucci}, {Oliva}, \&
  {Sturm}}]{maiolino2017}
{Maiolino}, R., {Russell}, H.~R., {Fabian}, A.~C., {et~al.} 2017, \nat, 544,
  202

\bibitem[{{McMullin} {et~al.}(2007){McMullin}, {Waters}, {Schiebel}, {Young},
  \& {Golap}}]{mcmullin2007}
{McMullin}, J.~P., {Waters}, B., {Schiebel}, D., {Young}, W., \& {Golap}, K.
  2007, in Astronomical Society of the Pacific Conference Series, Vol. 376,
  Astronomical Data Analysis Software and Systems XVI, ed. R.~A. {Shaw},
  F.~{Hill}, \& D.~J. {Bell}, 127

\bibitem[{{Noordermeer} {et~al.}(2007){Noordermeer}, {van der Hulst},
  {Sancisi}, {Swaters}, \& {van Albada}}]{noorder2007}
{Noordermeer}, E., {van der Hulst}, J.~M., {Sancisi}, R., {Swaters}, R.~S., \&
  {van Albada}, T.~S. 2007, \mnras, 376, 1513

\bibitem[{{Pearson} {et~al.}(2018){Pearson}, {Wang}, {Hurley}, {Ma{\l}ek},
  {Buat}, {Burgarella}, {Farrah}, {Oliver}, {Smith}, \& {van der
  Tak}}]{pearson2018}
{Pearson}, W.~J., {Wang}, L., {Hurley}, P.~D., {et~al.} 2018, \aap, 615, A146

\bibitem[{{Planck Collaboration} {et~al.}(2016){Planck Collaboration}, {Ade},
  {Aghanim}, {Arnaud}, {Ashdown}, {Aumont}, {Baccigalupi}, {Banday},
  {Barreiro}, {Bartlett}, {Bartolo}, {Battaner}, {Battye}, {Benabed},
  {Beno{\^\i}t}, {Benoit-L{\'e}vy}, {Bernard}, {Bersanelli}, {Bielewicz},
  {Bock}, {Bonaldi}, {Bonavera}, {Bond}, {Borrill}, {Bouchet}, {Boulanger},
  {Bucher}, {Burigana}, {Butler}, {Calabrese}, {Cardoso}, {Catalano},
  {Challinor}, {Chamballu}, {Chary}, {Chiang}, {Chluba}, {Christensen},
  {Church}, {Clements}, {Colombi}, {Colombo}, {Combet}, {Coulais}, {Crill},
  {Curto}, {Cuttaia}, {Danese}, {Davies}, {Davis}, {de Bernardis}, {de Rosa},
  {de Zotti}, {Delabrouille}, {D{\'e}sert}, {Di Valentino}, {Dickinson},
  {Diego}, {Dolag}, {Dole}, {Donzelli}, {Dor{\'e}}, {Douspis}, {Ducout},
  {Dunkley}, {Dupac}, {Efstathiou}, {Elsner}, {En{\ss}lin}, {Eriksen},
  {Farhang}, {Fergusson}, {Finelli}, {Forni}, {Frailis}, {Fraisse},
  {Franceschi}, {Frejsel}, {Galeotta}, {Galli}, {Ganga}, {Gauthier}, {Gerbino},
  {Ghosh}, {Giard}, {Giraud-H{\'e}raud}, {Giusarma}, {Gjerl{\o}w},
  {Gonz{\'a}lez-Nuevo}, {G{\'o}rski}, {Gratton}, {Gregorio}, {Gruppuso},
  {Gudmundsson}, {Hamann}, {Hansen}, {Hanson}, {Harrison}, {Helou},
  {Henrot-Versill{\'e}}, {Hern{\'a}ndez-Monteagudo}, {Herranz}, {Hildebrandt},
  {Hivon}, {Hobson}, {Holmes}, {Hornstrup}, {Hovest}, {Huang}, {Huffenberger},
  {Hurier}, {Jaffe}, {Jaffe}, {Jones}, {Juvela}, {Keih{\"a}nen}, {Keskitalo},
  {Kisner}, {Kneissl}, {Knoche}, {Knox}, {Kunz}, {Kurki-Suonio}, {Lagache},
  {L{\"a}hteenm{\"a}ki}, {Lamarre}, {Lasenby}, {Lattanzi}, {Lawrence}, {Leahy},
  {Leonardi}, {Lesgourgues}, {Levrier}, {Lewis}, {Liguori}, {Lilje},
  {Linden-V{\o}rnle}, {L{\'o}pez-Caniego}, {Lubin}, {Mac{\'\i}as-P{\'e}rez},
  {Maggio}, {Maino}, {Mandolesi}, {Mangilli}, {Marchini}, {Maris}, {Martin},
  {Martinelli}, {Mart{\'\i}nez-Gonz{\'a}lez}, {Masi}, {Matarrese}, {McGehee},
  {Meinhold}, {Melchiorri}, {Melin}, {Mendes}, {Mennella}, {Migliaccio},
  {Millea}, {Mitra}, {Miville-Desch{\^e}nes}, {Moneti}, {Montier}, {Morgante},
  {Mortlock}, {Moss}, {Munshi}, {Murphy}, {Naselsky}, {Nati}, {Natoli},
  {Netterfield}, {N{\o}rgaard-Nielsen}, {Noviello}, {Novikov}, {Novikov},
  {Oxborrow}, {Paci}, {Pagano}, {Pajot}, {Paladini}, {Paoletti}, {Partridge},
  {Pasian}, {Patanchon}, {Pearson}, {Perdereau}, {Perotto}, {Perrotta},
  {Pettorino}, {Piacentini}, {Piat}, {Pierpaoli}, {Pietrobon}, {Plaszczynski},
  {Pointecouteau}, {Polenta}, {Popa}, {Pratt}, {Pr{\'e}zeau}, {Prunet},
  {Puget}, {Rachen}, {Reach}, {Rebolo}, {Reinecke}, {Remazeilles}, {Renault},
  {Renzi}, {Ristorcelli}, {Rocha}, {Rosset}, {Rossetti}, {Roudier},
  {Rouill{\'e} d'Orfeuil}, {Rowan-Robinson}, {Rubi{\~n}o-Mart{\'\i}n},
  {Rusholme}, {Said}, {Salvatelli}, {Salvati}, {Sandri}, {Santos},
  {Savelainen}, {Savini}, {Scott}, {Seiffert}, {Serra}, {Shellard}, {Spencer},
  {Spinelli}, {Stolyarov}, {Stompor}, {Sudiwala}, {Sunyaev}, {Sutton},
  {Suur-Uski}, {Sygnet}, {Tauber}, {Terenzi}, {Toffolatti}, {Tomasi},
  {Tristram}, {Trombetti}, {Tucci}, {Tuovinen}, {T{\"u}rler}, {Umana},
  {Valenziano}, {Valiviita}, {Van Tent}, {Vielva}, {Villa}, {Wade}, {Wandelt},
  {Wehus}, {White}, {White}, {Wilkinson}, {Yvon}, {Zacchei}, \&
  {Zonca}}]{planck2016}
{Planck Collaboration}, {Ade}, P.~A.~R., {Aghanim}, N., {et~al.} 2016, \aap,
  594, A13

\bibitem[{{Rizzo} {et~al.}(2021){Rizzo}, {Vegetti}, {Fraternali}, {Stacey}, \&
  {Powell}}]{rizzo2021}
{Rizzo}, F., {Vegetti}, S., {Fraternali}, F., {Stacey}, H.~R., \& {Powell}, D.
  2021, \mnras, 507, 3952

\bibitem[{{Rizzo} {et~al.}(2020){Rizzo}, {Vegetti}, {Powell}, {Fraternali},
  {McKean}, {Stacey}, \& {White}}]{rizzo2020}
{Rizzo}, F., {Vegetti}, S., {Powell}, D., {et~al.} 2020, \nat, 584, 201

\bibitem[{{Sahu} {et~al.}(2019){Sahu}, {Graham}, \& {Davis}}]{sahu2019}
{Sahu}, N., {Graham}, A.~W., \& {Davis}, B.~L. 2019, \apj, 876, 155

\bibitem[{{Sandage}(1990)}]{sandage1990}
{Sandage}, A. 1990, \jrasc, 84, 70

\bibitem[{{Shao} {et~al.}(2019){Shao}, {Wang}, {Carilli}, {Wagg}, {Walter},
  {Li}, {Fan}, {Jiang}, {Riechers}, {Bertoldi}, {Strauss}, {Cox}, {Omont}, \&
  {Menten}}]{shao2019}
{Shao}, Y., {Wang}, R., {Carilli}, C.~L., {et~al.} 2019, \apj, 876, 99

\bibitem[{{Shao} {et~al.}(2022){Shao}, {Wang}, {Weiss}, {Wagg}, {Carilli},
  {Strauss}, {Walter}, {Cox}, {Fan}, {Menten}, {Narayanan}, {Riechers},
  {Bertoldi}, {Omont}, \& {Jiang}}]{shao2022}
{Shao}, Y., {Wang}, R., {Weiss}, A., {et~al.} 2022, arXiv e-prints,
  arXiv:2210.11926

\bibitem[{{Terzi{\'c}} \& {Graham}(2005)}]{terzic2005}
{Terzi{\'c}}, B. \& {Graham}, A.~W. 2005, \mnras, 362, 197

\bibitem[{{Toomre}(1977)}]{toomre1977a}
{Toomre}, A. 1977, in Evolution of Galaxies and Stellar Populations, ed. B.~M.
  {Tinsley} \& D.~C. {Larson}, Richard B.~Gehret, 401

\bibitem[{{Tripodi} {et~al.}(2022){Tripodi}, {Feruglio, C.}, {Fiore, F.},
  {Bischetti, M.}, {D\'{}Odorico, V.}, {Carniani, S.}, {Cristiani, S.},
  {Gallerani, S.}, {Maiolino, R.}, {Marconi, A.}, {Pallottini, A.},
  {Piconcelli, E.}, {Vallini, L.}, \& {Zana, T.}}]{Tripodi2022}
{Tripodi}, R., {Feruglio, C.}, {Fiore, F.}, {et~al.} 2022, A\&A, 665, A107

\bibitem[{{Vestergaard} \& {Osmer}(2009)}]{vosmer2009}
{Vestergaard}, M. \& {Osmer}, P.~S. 2009, \apj, 699, 800

\bibitem[{{Vestergaard} \& {Peterson}(2006)}]{vestergard2006}
{Vestergaard}, M. \& {Peterson}, B.~M. 2006, \apj, 641, 689

\bibitem[{{Vogelaar} \& {Terlouw}(2001)}]{gipsy}
{Vogelaar}, M.~G.~R. \& {Terlouw}, J.~P. 2001, in Astronomical Society of the
  Pacific Conference Series, Vol. 238, Astronomical Data Analysis Software and
  Systems X, ed. J.~{Harnden}, F.~R., F.~A. {Primini}, \& H.~E. {Payne}, 358

\bibitem[{{Volonteri}(2012)}]{volonteri2012}
{Volonteri}, M. 2012, Science, 337, 544

\bibitem[{{Wang} {et~al.}(2013){Wang}, {Wagg}, {Carilli}, {Walter}, {Lentati},
  {Fan}, {Riechers}, {Bertoldi}, {Narayanan}, {Strauss}, {Cox}, {Omont},
  {Menten}, {Knudsen}, {Neri}, \& {Jiang}}]{wang2013}
{Wang}, R., {Wagg}, J., {Carilli}, C.~L., {et~al.} 2013, \apj, 773, 44

\bibitem[{{Warner} {et~al.}(1973){Warner}, {Wright}, \& {Baldwin}}]{warner1973}
{Warner}, P.~J., {Wright}, M.~C.~H., \& {Baldwin}, J.~E. 1973, \mnras, 163, 163

\end{thebibliography}

\begin{appendix} 
        %
        \section{Details of the mass models}
        \label{sec:models-details}
        
        In this section, we describe the mass models we adopted in more detail. The gravitational contribution of the gas disc was computed using the task \texttt{Rotmod} in the \textsc{Gipsy} software \citep{gipsy}. It numerically solves the equation from \citet{casertano1983} for a truncated disc of finite thickness. For the radial column density distribution, we used the observed [CII] surface brightness profile as input, which is close to exponential. For the vertical density distribution, we assumed an exponential profile with a scale height $z_0=160$ pc.
            This is motivated by the scaling relation between scale length and scale height that is observed in edge-on disc galaxies at $z=0$ \citep[e.g.][]{bershady2010}, but the precise value of $z_{0, \rm gas}$ has a very minor effect on our results. In the limiting case of a razor-thin exponential disc (i.e. $z_{0, \rm gas}=0$), our results would vary within 15\%, which is well within our current uncertainties on the fitted quantities. We did not use the CO(6-5) surface brightness emission \citep{Tripodi2022} because it was derived from data with lower resolution ($\sim 0.5$ arcsec; \citealt{feruglio2018}), therefore the observed CO surface brightness profile is more sparsely sampled than the [CII] profile. For J2310, precise determinations for $M_{\rm gas}$ were obtained using the CO(6-5) emission line together with the CO spectral line energy distribution \citep{feruglio2018, li2020, Tripodi2022}; we used the value of $4.4\times 10^{10}$\msun as a tight prior on $M_{\rm gas}$. This estimate was derived by fitting the spectral line energy distribution of high-J CO lines and assuming a CO-to-H$_2$ conversion factor $\alpha_{\rm CO}= 0.8$ \msun (K km s$^{-1}$ pc$^{-2}$)$^{-1}$, which is currently thought to be most appropriate for QSO-host galaxies \citep{carilli2013}. Systematic uncertainties on $\alpha_{\rm CO}$ in starburst AGN-host galaxies can be large even at $z\simeq0$ \citep{Bolatto2013}, but there is mounting evidence that their $\alpha_{\rm CO}$ are systematically smaller than those of Milky-Way-like galaxies \citep{accurso2017, pearson2018, lelli2022}. We associate a 25\% uncertainty with $M_{\rm gas}$. This tight prior  is necessary to break a degeneracy between $Y_{\rm gas}$ and $Y_{\rm *disc}$ that arises because gas disc and stellar disc provide velocity contributions with a similar shape.  The actual gas mass may be more uncertain than assumed here, but this assumption mostly impacts the relative masses of the stellar and gas discs but leaves our conclusions on the central mass concentration unaltered.

               The gravitational contribution of the star-forming stellar disc was computed in a similar way as for the gas disc. For the radial column density profile, we used the surface brightness profile from the ALMA sub-millimeter continuum map, which traces hot dust heated by young stars. The sub-millimeter emission can be used as a proxy for the distribution of the stellar disc because dust absorbs UV radiation from young stars and re-emits it at far-infrared wavelengths \citep{lelli2021}.  The continuum surface brightness profile in \citet{Tripodi2022} is well described by  an exponential profile with scale length $R_{\rm *disc}=0.71$ kpc, which further supports the use of dust emission as a proxy for the star-forming disc. For the distribution of the vertical surface density, we assumed an exponential profile with $z_{0}\simeq160$ pc as for the gas disc, but this assumption has a very minor effect on our results.
               
               The stellar bulge was modelled as a spherical projected Sérsic profile, so that its gravitational contribution is given by \citep{lima1999, terzic2005}
                \begin{equation}
                        V_{\rm bulge}(R) = \sqrt{\dfrac{{\rm G}M_{\rm bulge}}{R}\dfrac{\gamma(n(3-p),b(R/R_e)^{1/n})}{\Gamma(n(3-p))}},
                \end{equation}
                \noindent where $M_{\rm bulge}$ is the stellar bulge mass, $R_e$ is the effective radius, $n$ is the Sérsic index, $b=2n-1/3+0.009876/n$ for $0.5<n<10$, $\gamma$ and $\Gamma$ are the incomplete and complete gamma function, respectively, and $p=1.0 - 0.6097/n+0.05563/n^2$ for $0.6<n<10$ and $10^{-2}\le R/R_e \le 10^3$. We fixed $n=4$ (a De Vaucouleurs profile) and set $R_e=0.07$ arcsec $\sim$409 pc, that is, $\sim {\rm beam\ size}/2$. Lower values of $R_e$ would not significantly change the contribution of the bulge in the centre, making it very similar to a point source. In principle, $R_e$ could be determined from high-resolution rest-frame optical and/or near-infrared images after subtracting the QSO contribution, but data like this are currently not available. Alternatively, $R_e$ may be set as a free parameter in our model, but we do not have enough data points in the rotation curve to fit more than four parameters.
                
                The SMBH gives the velocity contribution of a point source,
                \begin{equation}
                        \label{eq:vbh}
                        V_{\rm BH}(R) = \sqrt{\dfrac{{\rm G}M_{\rm BH}}{R}},
                \end{equation}
                where $M_{\rm BH}$ is the black mass. In our models, $M_{\rm BH}$ could be either a free or a fixed parameter. In the latter case, it was fixed to $M_{\rm BH}=5.0\times 10^{9}\ \rm M_{\odot}$, derived from the MgII emission line profile (Mazzucchelli in prep., \citealt{Tripodi2022}).

         In our modelling, we did not consider the contribution of a dark matter (DM) component for several reasons: (1) in general, the DM contribution in massive galaxies becomes important in the outer regions, beyond the stellar disc, because it has a rising trend with the radius. This component would therefore not affect our main results about the central mass component. (2) In our specific case, the rotation curve was measured out to a radius of just 1.7 kpc, within which the gravitational potential is likely dominated by baryons \citep[in analogy to star-forming galaxies at slightly lower redshifts; e.g.][]{genzel2017, genzel2020, rizzo2020, rizzo2021, lelli2021}. (3) Our four data points do not enable us to fit for additional free parameters, which would be required if we were to add a DM component. Therefore, we opted for the simplest physically motivated mass model with the smallest number of free parameters.
        
        \section{Simulated observation}
        \label{app:simul}
        
        \begin{figure*}
                \centering
                \includegraphics[width=0.9\linewidth]{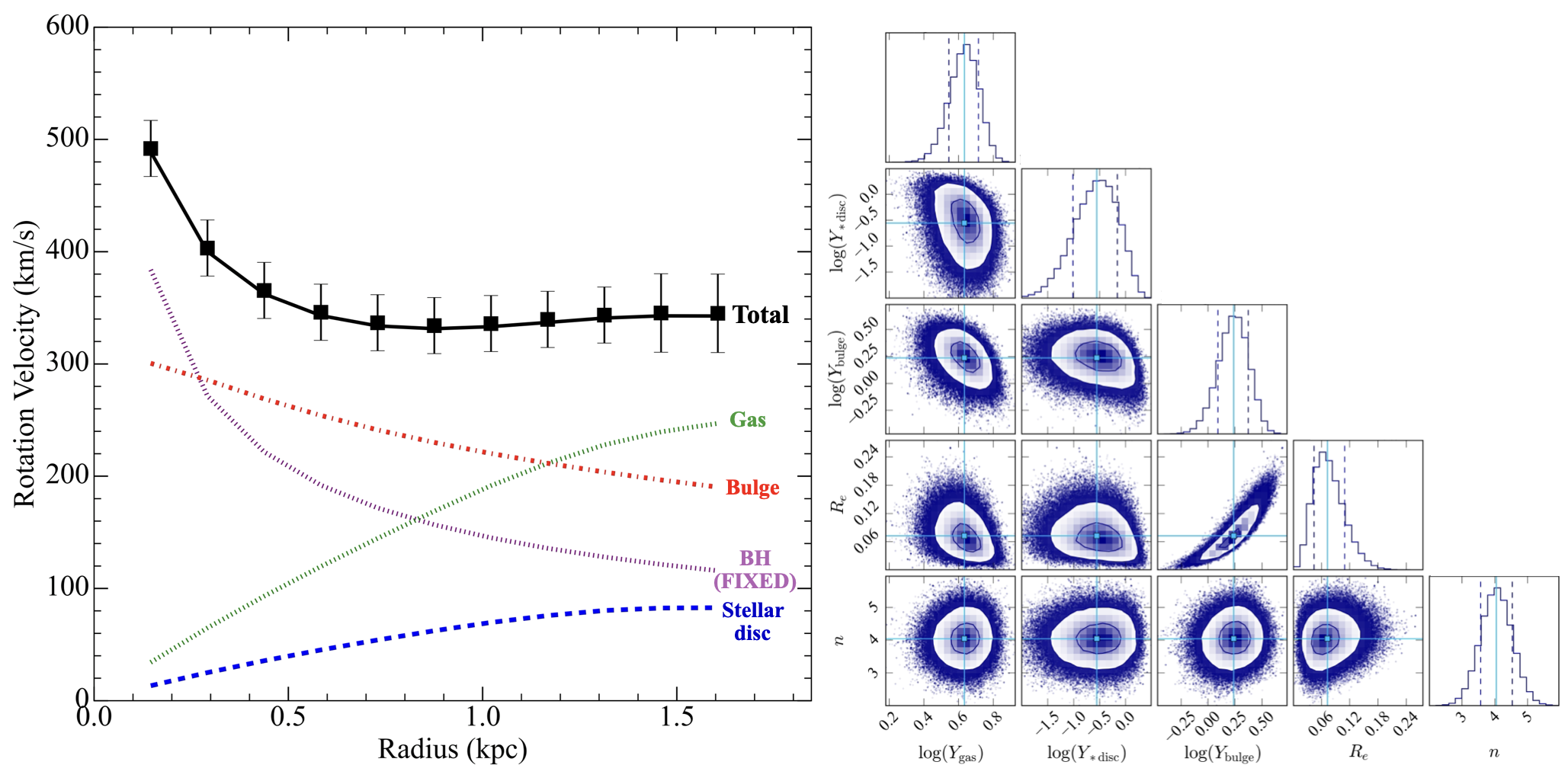}
                \caption{Results for the rotation curve of the simulated observation. Left: Best-fitting model for the simulated rotation curve. The best-fitting model (solid black line) is the sum of the contributions from the stellar disc (dashed blue line), the cold gas disc (dotted green line), the stellar bulge (dot-dashed red line), and the BH (dotted purple line). Right: Corner plot showing the four-dimensional posterior probability distributions of $\log(Y_{\rm gas}),\log(Y_{\rm *disc}),\log(Y_{\rm bulge}), R_e, n$. Blue lines indicate the best-fitting parameter, and the dashed lines mark the 16\% and 84\% percentiles for each parameter.}
                \label{fig:sim-models}
        \end{figure*}
        
        In this section, we evaluate the advantages of a very high resolution ALMA observation of the [CII] emission line in the QSO J2310. The goal of this observation would be to obtain an accurately sampled rotation curve, derived from a dynamical analysis of the [CII] emission line. This would allow us to quantify the contribution of the bulge with higher precision. 
        
        The parameters that describe the behaviour of a bulge in the rotation curve are the bulge mass, $M_{\rm bulge}$, the Sérsic index, $n$, and the effective radius, $R_{e}$ (see Sect. \ref{sec:model}). In order to constrain these three parameters, we would require a resolution that is of the order of $\sim R_{e}/2$. For our QSO at $z\sim6,$ we adopted $R_e=0.07$ arcsec (see Sect. \ref{sec:model}), therefore the required resolution is 0.05 arcsec, that is, three times lower than the resolution of the real observation we analysed in this work. First, we simulated a rotation curve that we would obtain using an observation with a resolution of 0.05 arcsec and given the results obtained in Sect. \ref{sec:model}. Secondly, we fit this simulated rotation curve using the mass models described in Appendix \ref{app:simul}, in order to assess the improvement achieved with higher-resolution data. 
        
        In Sect. \ref{sec:model} and Sect. \ref{sec:disc}, we concluded that the best model able to describe the shape of our observed rotation curve is model B, considering the contributions of a gaseous disc, a stellar disc, a BH, and a bulge. In order to build the simulated rotation curve, we fixed the mass of the components and the inclination at the values reported in Table \ref{tab:results}, and also $R_e=0.07, \ n=4$. We considered two points per resolution element and therefore obtained 11 simulated data points equally spaced by 0.025 arcsec (see the black squares in the left panel of Figure \ref{fig:sim-models}). We fit the simulated data set running an MCMC with five free parameters, $Y_{\rm gas}, Y_{\rm *disc}, Y_{\rm bulge}, R_{e}, n$, and a fixed BH mass ($M_{\rm BH}=5\times 10^9\ \rm M_{\odot}$). We adopted lognormal priors for the dimensionless mass parameters and Gaussian priors for $R_e$ and $n$. In particular, we used a tight prior on the gas mass ($\log(Y_{\rm gas})^{\rm prior}=0.64\pm 0.11$), and the priors on $\log(Y_{\rm *disc}), \log(Y_{\rm bulge}), R_{e}, n$ were centred near the values obtained (or fixed) for model B (see Table \ref{tab:results}) with standard deviations of $0.5,0.5,0.05,\text{and }0.5$ . The best-fitting model is shown in the left panel of Figure \ref{fig:sim-models} as a solid black curve together with its four components (gas in green, stellar disc in blue, BH in purple, and bulge in red). The corner plot and the numerical results are presented in the right panel of Figure \ref{fig:sim-models} and Table \ref{tab:sim-results}. The values of the best-fitting parameters agree very well with the expected values of model B, and the uncertainties are decreased from 20\% to 10\% for the gas mass and from 70\% to 30\% for the bulge mass. The uncertainty on the stellar disc mass is still high because this contribution has a strong degeneracy with the gas contribution, and therefore an independent measurement of the stellar disc mass would be required in order to place a tighter prior on $Y_{\rm *disc}$ and reduce its uncertainty. Similarly, a better estimation of $R_e$ would be obtained by a direct measurement of the bulge, for instance with the JWST. Nevertheless, the precision on the estimate of the bulge mass that is achieved with a very high resolution ALMA observation is four times higher than the precision obtained with our real observation. This result supports the request for a very high resolution (0.05 arcsec) ALMA observation, which has been proven to be able to place a tight constraint on the bulge mass.

        \begin{table}
                \caption{MCMC results for simulated rotation curve}
                \centering
                \begin{tabular}{c|ccccc}
                        \hline
                        Parameter & Best-fit   \\
                        \hline
                        $M_{\rm gas} [10^{10}\ \rm M_{\odot}]$ & $4.3\pm 0.6$\\[0.1cm]
                        $M_{\rm *disc} [10^{10}\ \rm M_{\odot}]$ &  $0.28\pm 0.27$\\[0.1cm]
                        $M_{\rm bulge} [10^{10}\ \rm M_{\odot}]$ &  $1.73\pm 0.53$\\[0.1cm]
                        $M_{\rm BH} [10^{10}\ \rm M_{\odot}]$ &  0.5 (Fixed) \\ [0.1cm]
                        $R_e$ [arcsec] & $0.07 \pm 0.04$\\
                        $n$ & $4.05\pm 0.48$\\
                        \hline
                \end{tabular}
                \label{tab:sim-results}
                \flushleft 

        \end{table}

\end{appendix}

\end{document}